\documentclass[prd,preprint,tightenlines,floatfix,showpacs,preprintnumbers,nofootinbib,eqsecnum,superscriptaddress]{revtex4}

 \usepackage[dvips,final]{graphicx}
  \usepackage{amssymb}
   \usepackage{amsmath}
    \usepackage{amsfonts}
     \usepackage{epsfig}
      \usepackage{bm}

\begin{document}

\title{Parton transverse momenta \\ and Drell-Yan
dilepton production}

\author{A.~Szczurek}
\email{Antoni.Szczurek@ifj.edu.pl}
\affiliation{Institute of Nuclear Physics PAN, PL-31-342 Cracow,
Poland} 
\affiliation{University of Rzesz\'ow, PL-35-959 Rzesz\'ow,
Poland}
\author{G. \'Slipek}
\email{Gabriela.Slipek@ifj.edu.pl}
\affiliation{Institute of Nuclear Physics PAN, PL-31-342 Cracow,
Poland} 

\date{\today}

\begin{abstract}
The differential cross section for the dilepton production
is calculated including Fermi motion of hadron
constituents as well as emission from the ladders in 
the formalism of unintegrated parton distributions. 
We use unintegrated parton distributions which fulfil 
Kwieci\'nski evolution equations.
Both zeroth- and first-order (for matrix element) 
contributions are included. 
We calculate azimuthal angular correlations between 
charged leptons
and deviations from the $p_t(l^+) = p_t(l^-)$ relation.
We concentrate on the distribution in dilepton-pair 
transverse momentum.
We find incident energy and virtuality dependence of
the distribution in transverse momentum of the lepton pair.
We study also azimuthal correlations between jet and dilepton pair
and correlation in the $(p_{1t}(jet),p_{2t}(l^+ l^-))$ space.
The results are compared with experimental data of 
the R209 and UA1 collaborations.
\end{abstract}


\maketitle

\section{Introduction}

The Drell-Yan dilepton production is one
of representative examples for which QCD collinear 
perturbative calculation can be performed order-by-order.
Usually inclusive distributions are discussed.
The most often studied observables are:
$d \sigma / dM_{ee}$ or $d \sigma /d x_F$, where
$M_{ee}$ is invariant mass of the dilepton pair
and $x_F$ is the Feynman variable of the pair.
In this paper we concentrate rather on one- and
two-dimensional distributions which are singular
in the standard collinear approximation.

In the 0-th order collinear approximation 
(quark-antiquark annihilation) the transverse momentum 
of the dilepton pair (sum of transverse momenta of opposite
sign charged leptons) is zero due to momentum conservation.
Then the 0th-order result is not included in calculating
the distribution in dilepton transverse momentum.
The lowest nonzero contributions are 1-st order 
quark-antiquark annihilation and QCD Compton.
Typical for collinear approach they show singularity
at small dilepton transverse momentum.

\begin{figure}[!ht]    %
\begin{center}
\includegraphics[width=0.6\textwidth]{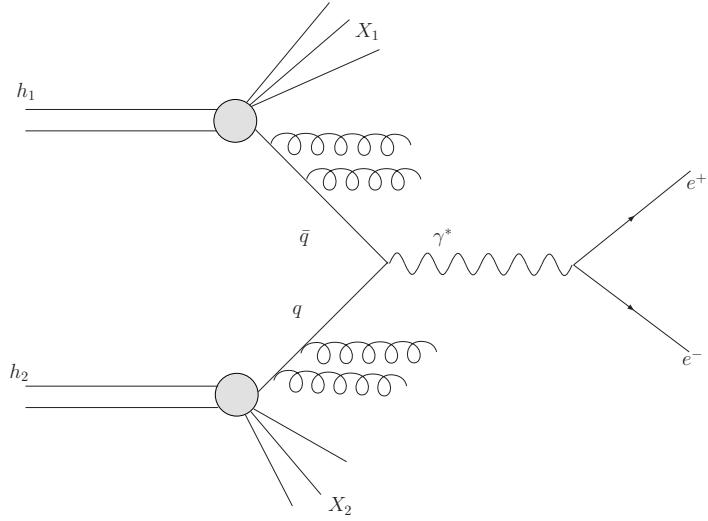}
\end{center}
   \caption{\label{fig:drell_yan_0th_order}
   \small
The diagram for the 0-th order Drell-Yan dilepton prodution
with initial emissions from the ladders.   
}
\end{figure}

Due to inter-quark interactions the quarks/antiquarks,
constituents of hadrons, are not at rest and posses
nonzero transverse momenta. Already this effect causes
that the 0-th order process contributes to the
finite transverse momenta of the lepton pair. Furthermore
the emissions of gluons before the $q \bar q \to l^+ l^-$
hard process causes an extra $k_t$-smearing which,
via momentum conservation, lead to finite transverse momenta
of the dilepton pair (see Fig.\ref{fig:drell_yan_0th_order}).
The initial transverse momenta are often modelled effectively
in terms of phenomenological Gaussian distributions \cite{WW98,AM04}.
The effect of Fermi motion as well as emission from the ladders
can be easily included in the formalism of Kwiecinski unintegrated 
parton distributions \cite{Kwiecinski}.

In the present paper we wish to calculate differential
cross section for dilepton production in the formalism of 
unintegrated parton distributions. We shall include both 
0-th order and 1-st order contributions.
We shall concentrate on the distributions in dilepton 
transverse momentum. This observable is extremely 
sensitive in the 0-th order to the initial transverse momenta 
of partons. The transverse momentum of Drell-Yan pair was
calculated within next-to-leading order perturbative QCD \cite{BGK98}
as well as in the resummation formalism in the impact parameter
space \cite{FQZ03}. Our approach differs in details from
those approaches. 

Our results will be compared with experimental data for
elementary proton-proton or proton-antiproton scattering.
We leave analysis of proton-nucleus scattering for a separate
publication.

\section{Formalism}

\subsection{0-th order Drell-Yan cross section}


The differential cross section for the 0-th order contribution
can be written as:
\begin{equation}
\begin{split}
\frac{d \sigma}{d y_1 d y_2 d^2p_{1t} d^2p_{2t}} = \sum_{f} \;
\int \frac{d^2 \kappa_{1t}}{\pi} \frac{d^2 \kappa_{2t}}{\pi}
\frac{1}{16 \pi^2 (x_1 x_2 s)^2} \; \\
\delta^2 \left( \vec{\kappa}_{1t} + \vec{\kappa}_{2t}
                 - \vec{p}_{1t} - \vec{p}_{2t} \right) \; 
[{\cal F}_{q_f}(x_1,\kappa_{1t}^2,\mu_F^2) \; {\cal F}_{\bar q_f}(x_2,\kappa_{2t}^2,\mu_F^2)\;
\overline{|M({q \bar q} \to {e^+ e^-})|^2 } \; \\
+ {\cal F}_{\bar q_f}(x_1,\kappa_{1t}^2,\mu_F^2) \; {\cal F}_{q_f}(x_2,\kappa_{2t}^2,\mu_F^2) \;
\overline{|M({q \bar q} \to {e^+ e^-})|^2 } \; ] \; ,
\end{split}
\label{0th_order_kt-factorization}
\end{equation}
where
${\cal F}_i(x_1,\kappa_{1t}^2)$ and ${\cal F}_i(x_2,\kappa_{2t}^2)$
are unintegrated quark/antiquark distributions in hadron $h_1$ and $h_2$,
respectively.

The longitudinal momentum fractions are evaluated in terms of
final lepton rapidities and transverse momenta:
\begin{eqnarray}
x_1 &=& \frac{m_{1t}}{\sqrt{s}}\exp( y_1) +\frac{m_{2t}}{\sqrt{s}}\exp( y_2) \;,\nonumber \\
x_2 &=& \frac{m_{1t}}{\sqrt{s}}\exp(-y_1) +\frac{m_{2t}}{\sqrt{s}}\exp(-y_2) ,
\end{eqnarray}
where $m_t = \sqrt{{p_t}^2 + m^2}$ is a so-called transverse mass.

The delta function in Eq.(\ref{0th_order_kt-factorization}) can be 
eliminated as e.g. in Refs.\cite{LS06,SRS07,PS07}.

Formally, if the following replacements
\begin{equation}
\begin{split}
&{\cal F}_i(x_1,\kappa_{1t}^2,\mu_F^2) \to x_1p_i(x_1,\mu_F^2) \delta(\kappa_{1t}^2) \; , \\
&{\cal F}_j(x_2,\kappa_{2t}^2,\mu_F^2) \to x_2p_j(x_2,\mu_F^2) \delta(\kappa_{2t}^2) \; 
\end{split}
\end{equation}
are done one recovers standard text book formulae.

\subsection{1-st order Drell-Yan cross section}

\begin{center}
\begin{figure*}    %
    \includegraphics[width=4.8cm]{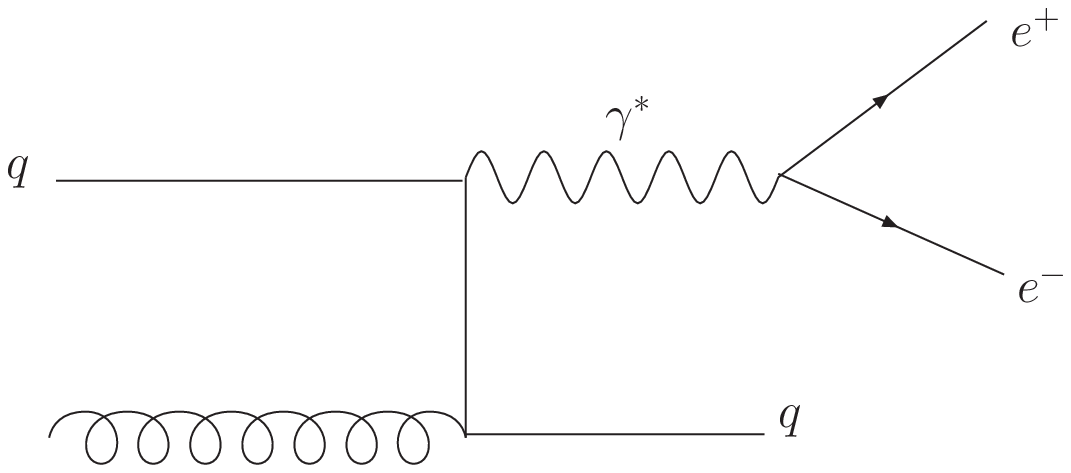}
    \includegraphics[width=4.8cm]{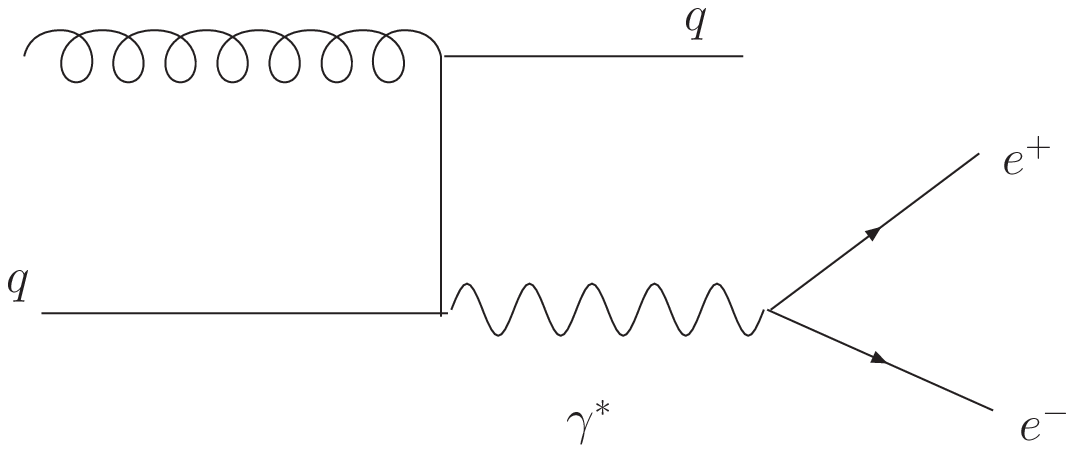} \\
    \includegraphics[width=4.8cm]{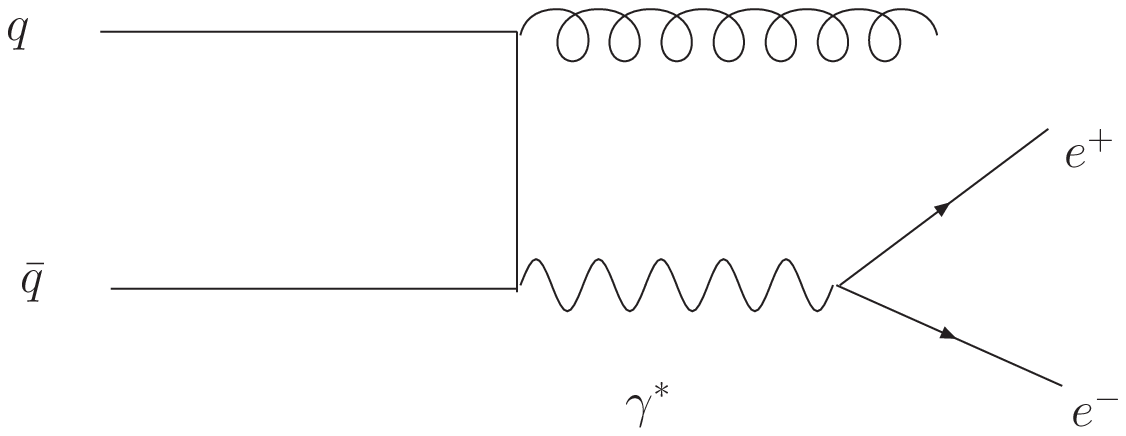}
    \includegraphics[width=4.8cm]{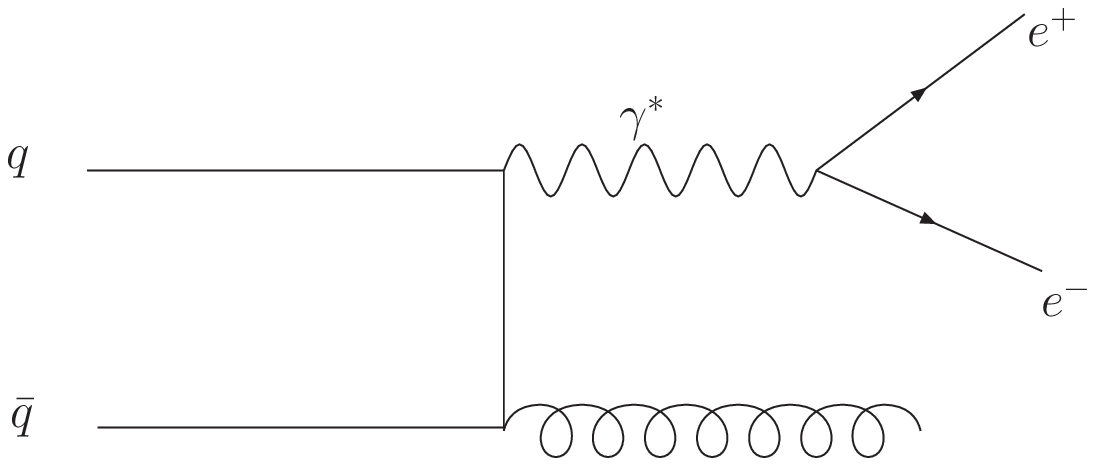}
   \caption{\label{fig:drell_yan_1st_order}
   \small
The subprocess diagrams for the 1-st order Drell-Yan 
dilepton prodution with initial emissions from the ladders.   
}
\end{figure*}
\end{center}

In the first order in $\alpha_s$ there are two types
of diagrams: QCD Compton and quark-antiquark annihilation.
A typical diagrams for corresponding subprocesses are shown 
in Fig.\ref{fig:drell_yan_1st_order}.

For example the multi-differential cross section for the QCD
Compton can be written in terms of unintegrated
quark/antiquark and gluon distributions as:
\begin{equation}
\begin{split}
\frac{d \sigma(h_1 h_2 \to \gamma^* X)}{d y_1 d y_2 d^2p_{1t} d^2p_{2t}} = \sum_{f} \;
\int \frac{d^2 \kappa_{1t}}{\pi} \frac{d^2 \kappa_{2t}}{\pi}
\frac{1}{16 \pi^2 (x_1 x_2 s)^2} \; \\
\delta^2 \left( \vec{\kappa}_{1t} + \vec{\kappa}_{2t}
                 - \vec{p}_{1t} - \vec{p}_{2t} \right) \; 
[ {\cal F}_g(x_1,\kappa_{1t}^2,\mu_F^2) \; {\cal F}_{q_f}(x_2,\kappa_{2t}^2,\mu_F^2)\;
\overline{|M({g q} \to {\gamma^* q})|^2 } \; \\
+ {\cal F}_{q_f}(x_1,\kappa_{1t}^2,\mu_F^2) \; {\cal F}_g(x_2,\kappa_{2t}^2,\mu_F^2) \;
\overline{|M({q g} \to {\gamma^* q})|^2 } \; ]  \; .
\end{split}
\label{1st_order_kt-factorization_Compton}
\end{equation}
Similarly the cross section for the first-order quark-antiquark annihilation associated with gluon emission can be written as
\begin{equation}
\begin{split}
\frac{d \sigma(h_1 h_2 \to \gamma^* X)}{d y_1 d y_2 d^2p_{1t} d^2p_{2t}} = \sum_{f} \;
\int \frac{d^2 \kappa_{1t}}{\pi} \frac{d^2 \kappa_{2t}}{\pi}
\frac{1}{16 \pi^2 (x_1 x_2 s)^2} \; \\
\delta^2 \left( \vec{\kappa}_{1t} + \vec{\kappa}_{2t}
                 - \vec{p}_{1t} - \vec{p}_{2t} \right) \; 
[ {\cal F}_{\overline{q}_f}(x_1,\kappa_{1t}^2,\mu_F^2) \; {\cal F}_{q_f}(x_2,\kappa_{2t}^2,\mu_F^2)\;
\overline{|M({\bar q q} \to {\gamma^* g})|^2 } \; \\
+ {\cal F}_{q_f}(x_1,\kappa_{1t}^2,\mu_F^2) \; {\cal F}_{\overline{q}_f}(x_2,\kappa_{2t}^2,\mu_F^2) \;
\overline{|M({q \bar q} \to {\gamma^* g})|^2 } \; ]  \; .
\end{split}
\label{1st_order_kt-factorization_annihilation}
\end{equation}
The delta functions in Eq.(\ref{1st_order_kt-factorization_Compton})
and Eq.(\ref{1st_order_kt-factorization_annihilation}) 
can be eliminated as e.g. in Refs.\cite{LS06,SRS07,PS07}.

The cross section for the emission of the dilepton pair 
can be expressed in terms of the cross section for 
the emission of time-like photon written above 
times a probability of the transition of the virtual 
photon into dilepton pair as:
\begin{equation}
\frac{d \sigma(h_1 h_2 \to {l^+ l^- j X})}{dM_{ll}^2}
 = 
\frac{\alpha_{em}}{3\pi M_{ll}^2}\;
d \sigma(h_1 h_2 \to \gamma^* j X) \; ,
\label{gamma*_ll}
\end{equation}
where $M_{ll}$ is the dilepton invariant mass.
Please note that $M_{ll}^2 = Q^2$, where $Q^2$ is 
virtuality of the time-like photon.

\subsection{Nonperturbative region of small $M_{ll}$}

The formalism presented up to here applies in the perturbative
region when the dilepton invariant mass $M_{ll}$ is not too small.
How to calculate Drell-Yan production for small invariant masses
($M_{ll} <$ 1 GeV) is not completely clear and this issue was not 
discussed in the literature.
In the region of very small photon virtualities 
the standard formulae presented
in the section above do not apply directly
and some modifications are necessary (the same is true for
soft-gluon resummation method).
The reason is twofold. First of all the parton distribution
may not exist for very small scales.
Secondly, there is a singularity when photon virtuality 
approaches 0.

In the present analysis we shall use the following extrapolation 
procedure:

Firstly, to be able to use the perturbative Kwieci\'nski 
parton distributions the factorization scale for UPDFs is taken as:
\begin{equation}
\mu_F^2 = M_{ll}^2 + Q_s^2 \; ,
\end{equation}
instead of $\mu_F^2 = M_{ll}^2$ used usually in calculating Drell-Yan
cross section for large dilepton invariant masses.

Secondly, to avoid singularities inherent in the matrix element
($\hat s \to$ 0) we use a simple replacement proposed in 
Ref.{\cite{wang00,AM04}:
\begin{equation}
\begin{split}
&\hat s \to \hat s + Q_s^2   \; , \\
&\hat t \to \hat t - Q_s^2/2 \; , \\
&\hat u \to \hat u - Q_s^2/2 \; .
\end{split}
\end{equation}
The parameter $Q_s^2$ is to be adjusted to experimental data.
The above procedure allows to avoid singularity when $M_{ll} \to$ 0.
Please note that such a replacement does not change 
$\hat s + \hat t + \hat u$.

The procedure described above is very similar in spirit to that used
in calculating the "deep inelastic" structure function $F_2$ 
for small photon virtualities 
\cite{Badelek-Kwiecinski,Szczurek-Uleshchenko}.

For illustration in Fig.\ref{fig:low_Q2} we present distributions
in transverse momentum of one of the leptons: electron
or positron (left panel) and distributions
in transverse momentum of the pair of leptons (right 
panel) for different values of the parameter $Q_s^2$.
In this calculation we have used unintegrated parton distributions to be discussed in the next subsection.
The results depend somewhat on the value of the parameter.
In principle, the shift parameter $Q_s^2$ could be adjusted
to experimental data for extremely low dilepton
invariant masses. It is not clear a priori if such
a procedure can be successful.


\begin{figure}
\begin{center}
\includegraphics[width=6.0cm]{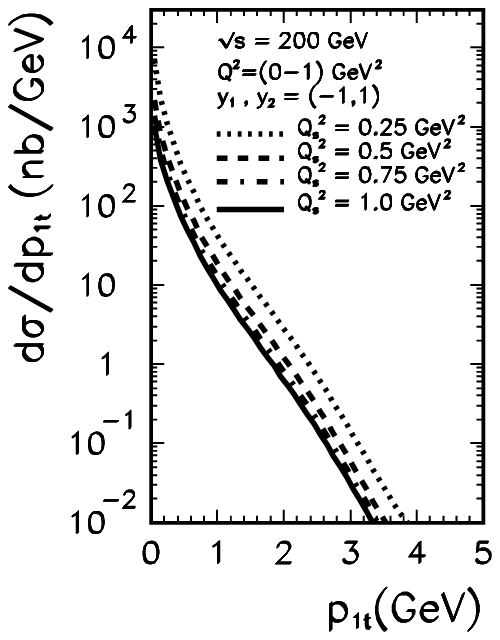}
\includegraphics[width=6.0cm]{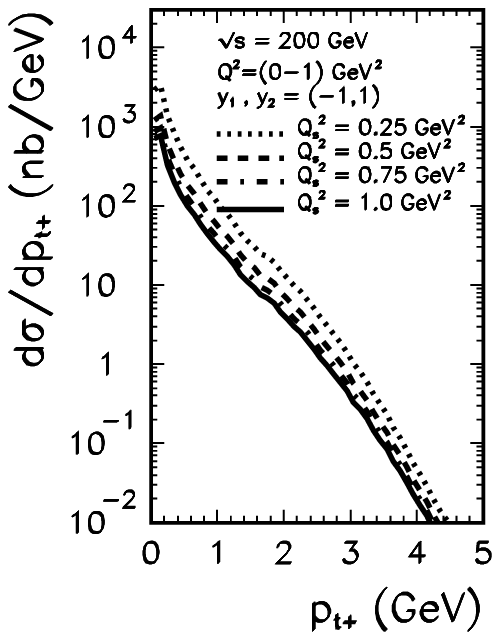}
\end{center}
\caption{\label{fig:low_Q2}
\small  
Distributions in transverse momentum
of the electron or positron (left panel)
and in the transverse momentum of the pair (right panel)
in the nonperturbative region of $Q^2 \in$ (0,1) GeV$^2$
at W = 200 GeV and for different values of the parameter
$Q_s^2$. In this calculation $Q_s^2$ = 0.25 (dotted), 0.5 (dashed), 0.75 (dash-dotted),
1.0 (solid) GeV$^2$ (from top to bottom).
Here midrapidity electrons and positrons 
$y_1, y_2 \in$ (-1,1) were selected for illustration.
}
\end{figure}

\subsection{Unintegrated parton distributions}

Due to its simplicity the Gaussian smearing of initial transverse momenta
is a good reference for other approaches. It allows to study
phenomenologically the role of transverse momenta in several
high-energy processes.
We define a simple unintegrated parton distributions:
\begin{equation}
{\cal F}_{i}^{Gauss}(x,\kappa^2,\mu_F^2) = x p_{i}^{coll}(x,\mu_F^2)
\cdot f_{Gauss}(\kappa^2) \; ,
\label{Gaussian_UPDFs}
\end{equation}
where $p_{i}^{coll}(x,\mu_F^2)$ are standard collinear (integrated)
parton distribution ($i = g, q, \bar q$) and $f_{Gauss}(\kappa^2)$
is a Gaussian two-dimensional function:
\begin{equation}
f_{Gauss}(\kappa^2) = \frac{1}{2 \pi \sigma_0^2}
\exp \left( -\kappa_t^2 / 2 \sigma_0^2 \right) / \pi \; .
\label{Gaussian}
\end{equation}
The UPDFs defined by Eq.(\ref{Gaussian_UPDFs}) and (\ref{Gaussian})
is normalized such that:
\begin{equation}
\int {\cal F}_{i}^{Gauss}(x,\kappa^2,\mu_F^2) \; d \kappa^2 = x
p_{i}^{coll}(x,\mu_F^2) \; .
\label{Gaussian_normalization}
\end{equation}
Kwieci\'nski has shown that the evolution equations
for unintegrated parton distributions takes a particularly
simple form in the variable conjugated to the parton transverse momentum.
In the impact-parameter space the Kwieci\'nski equation
takes the following simple form
\begin{equation}
\begin{split}
{\partial{\tilde f_{NS}(x,b,\mu^2)}\over \partial \mu^2} &=
{\alpha_s(\mu^2)\over 2\pi \mu^2}  \int_0^1dz  \, P_{qq}(z)
\bigg[\Theta(z-x)\,J_0((1-z) \mu b)\,
{\tilde f_{NS}\left({x\over z},b,\mu^2 \right)}
\\&- {\tilde f_{NS}(x,b,\mu^2)} \bigg]  \; , \\
{\partial{\tilde f_{S}(x,b,\mu^2)}\over \partial \mu^2} &=
{\alpha_s(\mu^2)\over 2\pi \mu^2} \int_0^1 dz
\bigg\{\Theta(z-x)\,J_0((1-z) \mu b)\bigg[P_{qq}(z)\,
 {\tilde f_{S}\left({x\over z},b,\mu^2 \right)}
\\&+ P_{qg}(z)\, {\tilde f_{G}\left({x\over z},b,\mu^2 \right)}\bigg]
 - [zP_{qq}(z)+zP_{gq}(z)]\,
{\tilde f_{S}(x,b,\mu^2)}\bigg\}  \; ,
 \\
{ \partial {\tilde f_{G}(x,b,\mu^2)}\over \partial \mu^2}&=
{\alpha_s(\mu^2)\over 2\pi \mu^2} \int_0^1 dz
\bigg\{\Theta(z-x)\,J_0((1-z) \mu b)\bigg[P_{gq}(z)\,
{\tilde f_{S}\left({x\over z},b,\mu^2 \right)}
\\&+ P_{gg}(z)\, {\tilde f_{G}\left({x\over z},b,\mu^2 \right)}\bigg]
-[zP_{gg}(z)+zP_{qg}(z)]\, {\tilde f_{G}(x,b,\mu^2)}\bigg\} \; .
\end{split}
\label{kwiecinski_equations}
\end{equation}
We have introduced here the short-hand notation
\begin{equation}
\begin{split}
\tilde f_{NS}&= \tilde f_u - \tilde f_{\bar u}, \;\;
                 \tilde f_d - \tilde f_{\bar d} \; ,  \\
\tilde f_{S}&= \tilde f_u + \tilde f_{\bar u} + 
                \tilde f_d + \tilde f_{\bar d} + 
                \tilde f_s + \tilde f_{\bar s} \; . 
\end{split}
\label{singlet_nonsinglet}
\end{equation}
The unintegrated parton distributions in the impact factor
representation are related to the familiar collinear distributions
as follows
\begin{equation}
\tilde f_{k}(x,b=0,\mu^2)=\frac{x}{2} p_k(x,\mu^2) \; .
\label{uPDF_coll_1}
\end{equation}
On the other hand, the transverse momentum dependent UPDFs are related
to the integrated parton distributions as
\begin{equation}
x p_k(x,\mu^2) =
\int_0^{\infty} d \kappa_t^2 \; {\cal F}_k(x,\kappa_t^2,\mu^2) \; .
\label{uPDF_coll_2}
\end{equation}

The two possible representations are interrelated via Fourier-Bessel
transform
\begin{equation}
  \begin{split}
    &{{\cal F}_k(x,\kappa_t^2,\mu^2)} =
    \int_{0}^{\infty} db \;  b J_0(\kappa_t b)
    {{\tilde f}_k(x,b,\mu^2)} \; ,
    \\
    &{{\tilde f}_k(x,b,\mu^2)} =
    \int_{0}^{\infty} d \kappa_t \;  \kappa_t J_0(\kappa_t b)
    {{\cal F}_k(x,\kappa_t^2,\mu^2)} \; .
  \end{split}
\label{Fourier}
\end{equation}
The index k above numerates either gluons (k=0), quarks (k$>$ 0) or
antiquarks (k$<$ 0).

While physically ${\cal F}_k(x,\kappa_t^2,\mu^2)$ should be positive,
there is no obvious reason for such a limitation for
$\tilde f_k(x,b,\mu^2)$.

In the following we use leading-order parton distributions
from Ref.\cite{GRV98} as the initial condition for QCD evolution.
The set of integro-differential equations in b-space
was solved by the method based on the discretisation made with
the help of the Chebyshev polynomials (see \cite{Kwiecinski}).
Then the unintegrated parton distributions were put on a grid
in $x$, $b$ and $\mu^2$ and the grid was used in practical
applications for Chebyshev interpolation. 

For the calculation of inclusive and coincidence cross section
for the heavy (time-like) photon production (see next section) 
the parton distributions in momentum space are more useful.
These calculation requires a time-consuming multi-dimensional
integration. An explicit calculation of the Kwieci\'nski UPDFs
via Fourier transform for needed in the main calculation values of
$(x_1,\kappa_{1t}^2)$ and $(x_2,\kappa_{2t}^2)$
is not possible.
Therefore it becomes a neccessity to prepare auxiliary grids of
the momentum-representation UPDFs before
the actual calculation of the cross sections.
These grids are then used via a two-dimensional interpolation
in the spaces $(x_1,\kappa_{1t}^2)$ and $(x_2,\kappa_{2t}^2)$
associated with each of the two incoming partons.

\section{A comment on $b$-space resummation}

There is an alternative approach to calculate
transverse momentum distribution of the dilepton pair.
In the Collins-Soper-Sterman formalism \cite{CSS} 
(see also\cite{FQZ03}), known
also as $b$-space resummation, the cross section
differential in dilepton invariant mass ($M_{ll}$),
rapidity of the pair ($y$) and transverse momentum
of the pair ($p_{t+}$) can be written as
\begin{equation}
\frac{d \sigma}{d M_{ll}^2 dy dp_t^2} =
\frac{1}{(2\pi)^2} \int d^2b 
\exp(i \vec{p}_t \vec{b}) W(b,M_{ll},x_1,x_2) =
\frac{1}{2 \pi} \int db b J_0(p_{t+} b) W(b,M_{ll},x_1,x_2) \; .
\label{soft_resummation}
\end{equation}
The integrand function $W(b,M_{ll}^2,p_{t+},x_1,x_2)$ for 
the Drell-Yan pair production is:
\begin{equation}
\begin{split}
&W(b,M_{ll}^2,p_{t+},x_1,x_2) = \frac{4 \pi^2 \alpha^2}{9 M_{ll}^2 s}
\sum_f e_f^2 \;
e^{-S(b,M_{ll}^2)} \\
&\left( \sum_a 
\int_{x_1}^1 \frac{d \xi_1}{\xi_1} 
f_{a/h_1} \left( \xi_1,\frac{1}{b^2} \right)
C_{fa} \left( \frac{x_1}{\xi_1};b \right) \right) \\
&\left( \sum_b
\int_{x_2}^1 \frac{d \xi_2}{\xi_2} 
f_{b/h_2} \left( \xi_1,\frac{1}{b^2} \right)
C_{fb} \left( \frac{x_2}{\xi_2};b \right) \right) \\
&+
\frac{4 \pi^2 \alpha^2}{9 M_{ll}^2 s} Y(M_{ll}^2,p_{t+},x_1,x_2) \; .
\end{split}
\label{W_function}
\end{equation}
The first term is naturally called resummation term
and the function $Y$ gives a correction which is 
negligible for small $p_{t+}$. The decomposition 
(\ref{W_function}) is not free of ambiguities, 
like matching condition etc. Furthermore it is not easy
to assure that in each corner of the phase space 
($y$, $M_{ll}$, $p_{t+}$) the resummation part of 
the cross section is positive. 
For $p_{t+} \ll M_{ll}$ the resummation term is much bigger
than the correction term.
The formula (\ref{soft_resummation}) together with
(\ref{W_function}) has a singularity when $M_{ll} \to$ 0
due to the photon propagator.
In Eq.(\ref{W_function}) the exponent in the Sudakov-like 
form factors reads
\begin{equation}
S(b,Q^2) = \int_{\mu_{min}^2(b)}^{Q^2} 
\frac{d \mu^2}{\mu^2} \;
\left[ 
\log \left( \frac{Q^2}{\mu^2} \right)
A(\alpha_s(\mu^2)) +  B(\alpha_s(\mu^2))                   
\right] \; .
\label{Sudakov_formfactor}
\end{equation}
The coefficients $A$ and $B$ can be expanded in the 
series of $\alpha_s$ \cite{CSS}.

In this approach longitudinal momentum fractions
are calculated as a rule as for the collinear 0th-order
kinematics:
\begin{eqnarray}
x_1 &=& \exp(+y) \frac{M_{ll}}{\sqrt{s}}, \nonumber \\
x_2 &=& \exp(-y) \frac{M_{ll}}{\sqrt{s}} \; .
\end{eqnarray}
Please note that this is slightly different than
in our case, where the transverse momenta of 
leptons explicitly enter into
corresponding formulae. The soft-gluon resummation
formula could be corrected for finite $p_{t+}$.

The coefficient functions $C_{fi}$ can be expanded 
in terms of $\alpha_s$
\begin{eqnarray}
C_{fa} &=& \delta_{fa} \delta(1-\xi_1) + ...   \; ,
\nonumber \\ 
C_{fb} &=& \delta_{fb} \delta(1-\xi_2) + ...   \; ,
\label{C_factors}
\end{eqnarray}
where the dots represent higher order terms in $\alpha_s$.
Limiting to the resummation term in (\ref{W_function}) 
and keeping only first terms in the expansion 
(a bit academic approximation used here for simplicity) 
one gets
\begin{equation}
\begin{split}
W^{res}(b,M_{ll}^2,p_t,x_1,x_2) \approx
\frac{4 \pi^2 \alpha^2}{9 M_{ll}^2 s} \sum_f \; e_f^2 
e^{-S(b,M_{ll}^2)} \\
\left[ 
q_f \left( x_1,\frac{1}{b^2} \right) 
{\bar q}_{f} \left( x_2,\frac{1}{b^2} \right) +
{\bar q}_{f} \left( x_1,\frac{1}{b^2} \right) 
q_f \left( x_2,\frac{1}{b^2} \right) 
\right] \; .
\end{split}
\label{simplified_formula}
\end{equation}
Obviously $b \ll 1/\Lambda$ in order that the pQCD is 
at work.
The region of large $b$ is of nonperturbative nature.
In order to supplement to the whole $b$ space an 
extrapolation is needed.
Usually this is done with the help of the following
prescription:
\begin{equation}
b \to b_* = \frac{b}{(1 + b^2/b_{max}^2)^{1/2}} < b \; .
\label{effective_b}
\end{equation}
Above $b_{max}$ is a free, a bit arbitrary parameter
($1/b_{max}^2 > \mu_0^2$, where $\mu_0^2$ is a minimal
possible factorization scale).
To cut off contributions of large impact factor an extra
form factor $F^{NP}(b,x_1,x_2,...)$ multiplying 
$W^{res}(b,M_{ll},p_{t+},x_1,x_2)$ is introduced. 
Above dots represent a set of free parameters.
These parameters must be found by fitting to experimental 
data. 
Typical, not small, uncertainties are shown e.g. 
in Fig.1 of Ref.\cite{FQZ03} (see also \cite{LY94}).

Our approach here was formulated in the momentum space.
Let us limit (as in the b-space resummation) to
distributions in $y$, $p_{t+}$ and $M_{ll}^2$.
Then the cross section can be written approximately(!)
as
\begin{equation}
\begin{split}
\frac{d \sigma}{d y d^2 p_{t+} dM_{ll}^2}
\approx \sum_f \frac{\sigma_f^{DY}}{s x_1 x_2} \int 
\frac{d^2 k_{1t}}{\pi}  \frac{d^2 k_{2t}}{\pi} \\
\left(
f_{q_f}(x_1,k_{1t}^2,\mu_{F}^2) f_{{\bar q}_f}(x_2,k_{2t}^2,\mu_{F}^2) +
f_{{\bar q}_f}(x_1,k_{1t}^2,\mu_{F}^2) f_{q_f}(x_2,k_{2t}^2,\mu_{F}^2)
\right)  
\; \delta^2 \left(
\vec{k}_{1t} + \vec{k}_{2t} - \vec{p}_{t+} \right)
\; ,
\end{split}
\end{equation}
where $\sigma_{f}^{DY} = \frac{1}{3} e_f^2 \frac{4 \pi \alpha^2}{3 \hat{s}}$.
Analogously as for the $W$ and $Z$ boson production
(see Ref.\cite{KS03}) the spectrum in $y$, $p_{t+}$
and $M_{ll}^2$ can be written also in terms of $b$-space
unintegrated parton distributions.
The corresponding cross section then reads:
\begin{equation}
\begin{split}
\frac{d \sigma}{d y d^2 p_{t+} dM_{ll}^2}
= \frac{4 \pi \alpha^2}{9 M_{ll}^2 {\hat s}} 
\frac{1}{\pi^2}
\sum_f e_f^2
\int d^2 b J_0(p_{t+} b) \\
\left[
  {\tilde f}_{q/1}(x_1,b,\mu_F^2) {\tilde f}_{\bar q/2}(x_2,b,\mu_F^2)
+ {\tilde f}_{\bar q/1}(x_1,b,\mu_F^2) {\tilde f}_{q/2}(x_2,b,\mu_F^2)
\right]
\; .
\end{split}
\label{b-space_kt_factorization}
\end{equation}
In most of the evalutions presented in this paper 
$\mu_F^2 = M_{ll}^2$ is used.

We wish to note very similar structure of the $b$-space 
resummation formula (\ref{soft_resummation}) with
(\ref{simplified_formula}) and our 
formula (\ref{b-space_kt_factorization}).
Asumming a factorizable form for the nonperturbative
form factor \footnote{This is justified if the main
nonperturbative effects are due to the merging of
initial (anti)quarks in bound (nonpertubative) hadrons
which causes that initial partons are not at rest
and posses internal momenta.}:
\begin{equation}
F_{q \bar q}^{NP}(Q,b,x_1,x_2) = 
F_{q}^{NP}(Q,b,x_1) \cdot F_{\bar q}^{NP}(Q,b,x_2) \; 
\end{equation}
one could define $b$-space resummation unintegrated
(anti)quark distributions in the b-space as:
\begin{equation}
\begin{split}
f_{{q}_f}^{SGR}(x_1,b,Q^2) = F_{q}(b,x_1,Q^2) 
\left[
x_1 q_f(x_1,\mu^2(b)) + ...
\right]
\exp \left( -\frac{1}{2} S(b,Q^2) \right) \; , \\
f_{{\bar q}_{f}}^{SGR}(x_2,b,Q^2) = F_{\bar q}(b,x_2,Q^2) 
\left[
x_2 {\bar q}_f(x_2,\mu^2(b)) + ...
\right]
\exp \left( -\frac{1}{2} S(b,Q^2) \right) \; .
\end{split}
\end{equation}

The $b$-space resummation is useful for calculating
transverse momentum distribution of the dilepton pair,
a variable Fourier-conjugated to $b$.
For other correlation observables between $e^+$ and $e^-$
exact kinematics and new kinematical
variables must be used.
This seems not possible in the framework of 
the $b$-space resummation.
Our approach, formulated in the momentum space and 
with explicit kinematics of each lepton, is better
suited for calculating observables like
$\frac{d \sigma}{d \phi_{ee}}$ or
$\frac{d \sigma}{dp_{t,e^+} d p_{t,e^-}}$ discussed in
the present paper.

\section{Results}

\subsection{0-th order component with transverse momenta}

In the standard collinear approach in 0-th order
approximation leptons of opposite charge are produced
back-to-back, i.e. the relative azimuthal angle between
them is fixed and equals to 180$^o$.
The situation changes if one includes "initial"
transverse momenta of partons which annihilate producing
dilepton pair. This is shown in Fig.\ref{fig:dsig_dphid_gau_kw}.


\begin{figure}
\begin{center}
\includegraphics[width=0.5\textwidth]{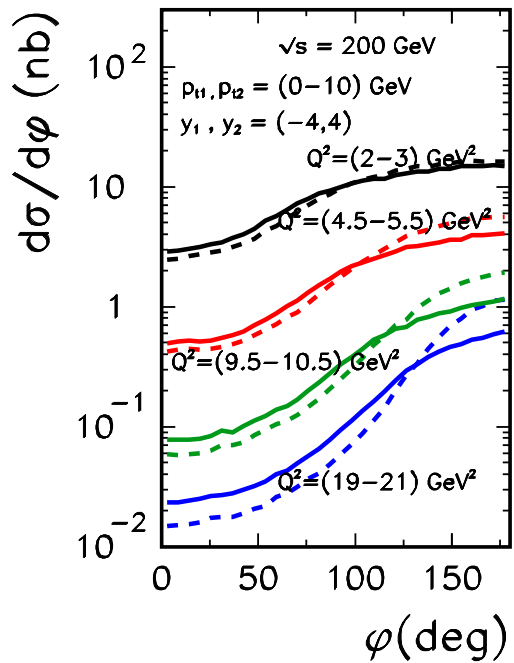}
\end{center}
   \caption{\label{fig:dsig_dphid_gau_kw}
   \small
Distributions in azimuthal angle between electron and 
positron in proton-proton scattering for RHIC energy 
$\sqrt{s}$ = 200 GeV and different windows in $Q^2$ 
specified in the figure.
The dashed lines are for UPDFs with Gaussian smearing 
($\sigma_0 = \frac{\sqrt{2}}{2} \frac{1}{b_0}$, $b_0$ = 1 GeV$^{-1}$)
and the solid lines are for Kwieci\'nski UPDFs with
$b_0$ = 1 GeV$^{-1}$. Here $\mu_F^2 = Q^2$ was chosen.
No cuts on lepton transverse momenta were applied.
}
\end{figure}


We have selected RHIC energy ($\sqrt{s}$ = 200 GeV) for illustration.
Since the plot is purely theoretical we did not make any cuts
on electron/positron transverse momenta and rapidities.
We show results for narrow bins in photon virtuality (= square of
the dilepton invariant mass) specified in the figure. 
Of course, the cross section strongly depends on photon virtuality; 
the smaller virtuality the bigger cross section. 
This is mainly a kinematical effect. This can be seen
by comparison of solid (Kwieci\'nski UPDFs)
and dashed (Gaussian smearing) lines.
In this calculation $\sigma_0$ of the Gaussian distribution
was adjusted to $b_0$ in the Kwieci\'nski UPDFs.
For small virtualities the two distributions almost coincide.
At large virtuality they differ much more which can be
understood as effect of evolution of UPDFs from the initial scale 
$\mu_0^2$ to $Q^2$.

How the azimuthal angle correlations depend on cuts in lepton
transverse momenta ?
In Fig.\ref{fig:dsig_dphid_pt_cuts} we show some examples for the E772
experiment at Fermilab. The cuts not only lower the cross section but also
modify the shape of azimuthal correlations.


\begin{figure}
\begin{center}
\includegraphics[width=0.5\textwidth]{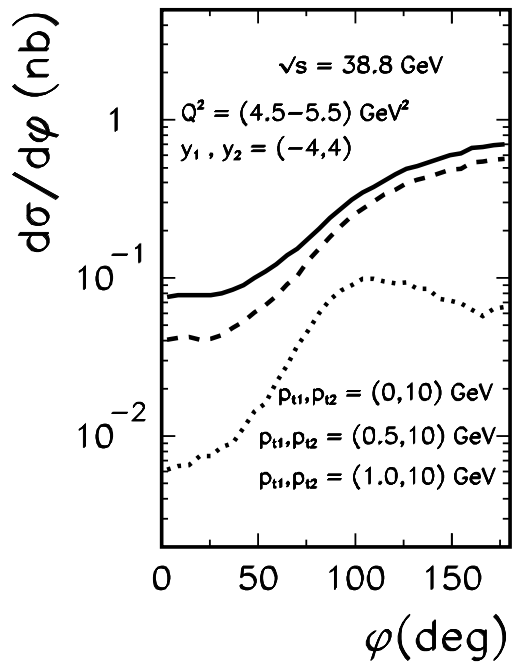}
\end{center}
   \caption{\label{fig:dsig_dphid_pt_cuts}
   \small
The influence of cuts on lepton transverse momenta on distributions 
in azimuthal angle between electron and positron 
for $\sqrt{s}$ = 38.8 GeV (E772 experiment).
Here $\mu_F^2 = Q^2$ was chosen.
}
\end{figure}


In collinear 0th-order approximation the electron
and positron transverse momenta compansate each other, i.e.
$p_{1t}(e^+) = p_{2t}(e^-)$. In the $k_t$-factorization approach
this condition is relaxed. As an example in 
Fig.\ref{fig:maps_p1tp2t} we show two dimensional maps
in the region of relatively small transverse momenta
($p_{1t},p_{2t} <$ 5 GeV). In this calculation we did
not put any constraint on photon virtuality (dimuon invariant mass).
In the left panel we show results obtained with 
Eq.(\ref{0th_order_kt-factorization}).
As discussed in the theoretical section real perturbative
calculation requires presence of large scales. Therefore
to be precise the perturbative calculation at small transverse
momenta is not reliable.
This can be better understood by looking at 
Fig.\ref{fig:map_aveq2_p1tp2t} which shows correlations of 
transverse momenta and photon virtualities.
In the right panel of Fig.\ref{fig:maps_p1tp2t} we show in addition
calculation with shifted scales and kinematical variables.
A careful inspection of both panels shows that the differences
are only at small transverse momenta. A more orthodox approach would be 
to completely exclude the region of small transverse momenta of leptons,
which means also small dilepton invariant masses.

\begin{figure}
\includegraphics[width=0.4\textwidth]{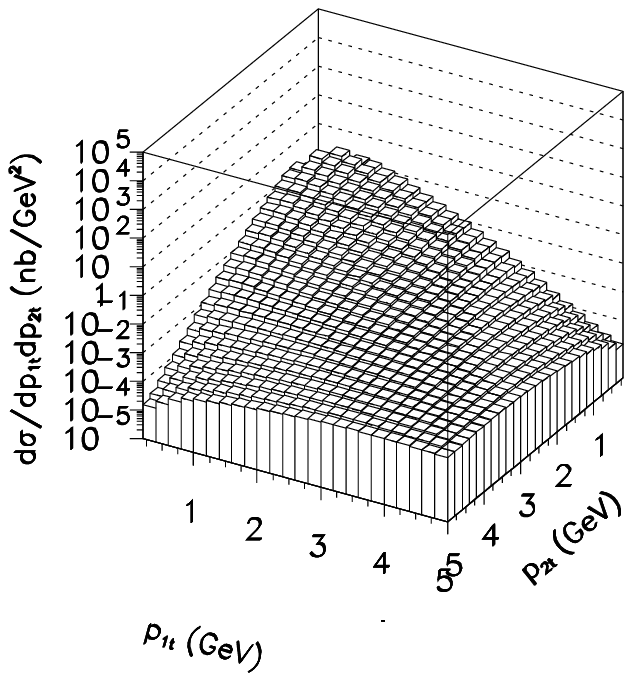}
\includegraphics[width=0.4\textwidth]{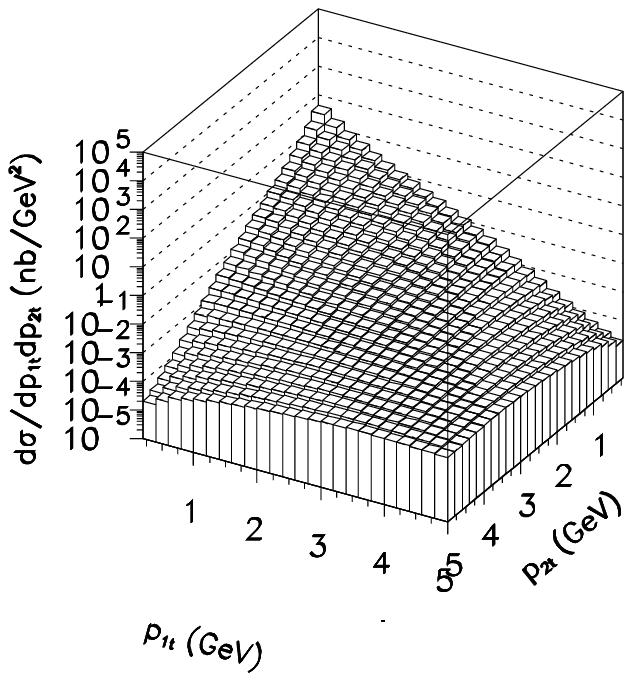}
   \caption{\label{fig:maps_p1tp2t}
   \small
Distribution in ($p_{1t}(e^+), p_{2t}(e^-)$)
for zeroth-order Drell-Yan in proton-proton collisions at 
$\sqrt{s}$ = 200 GeV. 
In the left panel standard procedure is used with $\mu_F^2 = Q^2$
and in the right panel in addition $Q_{s}^2$ = 1 GeV$^2$ is used 
as described in subsection IIC.
In this calculation -1 $< y_1, y_2 <$ 1.
}
\end{figure}


\begin{figure}
\begin{center}
\includegraphics[width=0.5\textwidth]{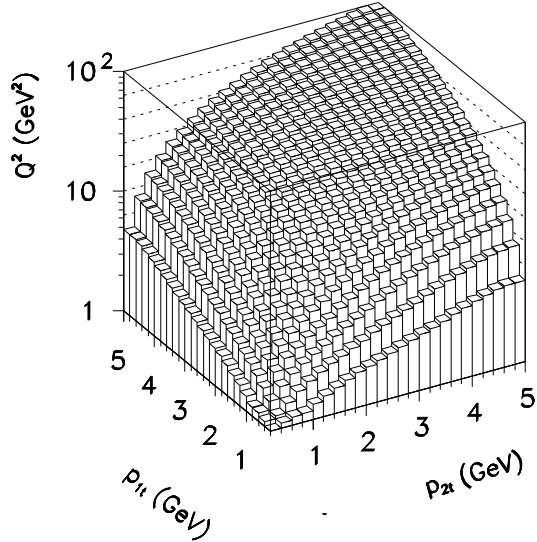}
\end{center}
   \caption{\label{fig:map_aveq2_p1tp2t}
   \small
Average value of the photon virtuality on the ($p_{1t}(e^+), p_{2t}(e^-)$)
plane for zeroth-order Drell-Yan in proton-proton collisions
at $\sqrt{s}$ = 200 GeV. 
Here $\mu_F^2 = Q^2$ and -1 $< y_1, y_2 <$ 1.   
}
\end{figure}


Very interesting observable, which is singular in collinear
approximation in leading-order, is the distribution in the transverse
momentum of the dilepton pair ($p_{t+}$). 
In Fig.\ref{fig:dsig_dptsum_diff_energies}
we show such a dependence on the incident center-of-mass energy
for two different bins in photon virtuality. In general,
the bigger energy the broder the distribution in $p_{t+}$
\footnote{In collinear approach the distribution would be delta 
function in transverse momentum of the pair.}. 
The bigger energies correspond to smaller values
of quark/antiquark longitudinal momentum fractions. The effect of
broadening of the distribution is larger for larger photon-virtuality
which is taken as a factorization scale in the Kwieci\'nski UPDFs.


\begin{figure}
\begin{center}
\includegraphics[width=0.4\textwidth]{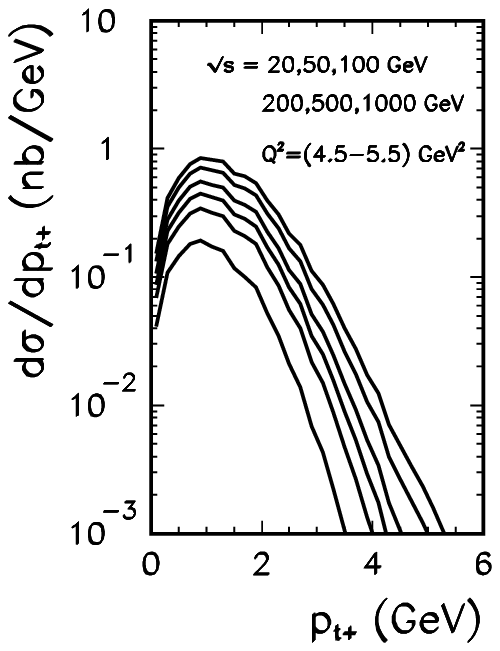}
\includegraphics[width=0.4\textwidth]{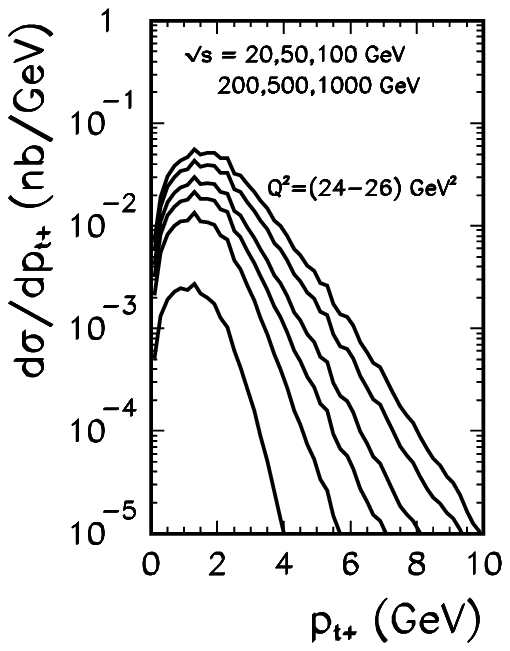}
\end{center}
   \caption{\label{fig:dsig_dptsum_diff_energies}
   \small
Distribution in $p_{t+}$ for different beam energies specified in the
figure and for two different windows in $Q^2$. Here $\mu_F^2 = Q^2$.  
}
\end{figure}


The effect of broadening is summarized in Fig.\ref{fig:pt_plus_ave}
where is show average value of the lepton-pair transverse momentum.
The difference between different bins in photon virtuality is
due to QCD evolution effect (we use photon virtuality $Q^2$ as
a factorization scale in the Kwieci\'nski unintegrated parton
distributions). Similar effects were already discussed
for dijet \cite{SRS07} and photon-jet correlations \cite{PS07}.
The two curves in Fig.\ref{fig:pt_plus_ave} would coincide if 
the evolution effect would be neglected, as was done e.g. in Ref.\cite{AM04}.


\begin{figure}
\begin{center}
\includegraphics[width=0.6\textwidth]{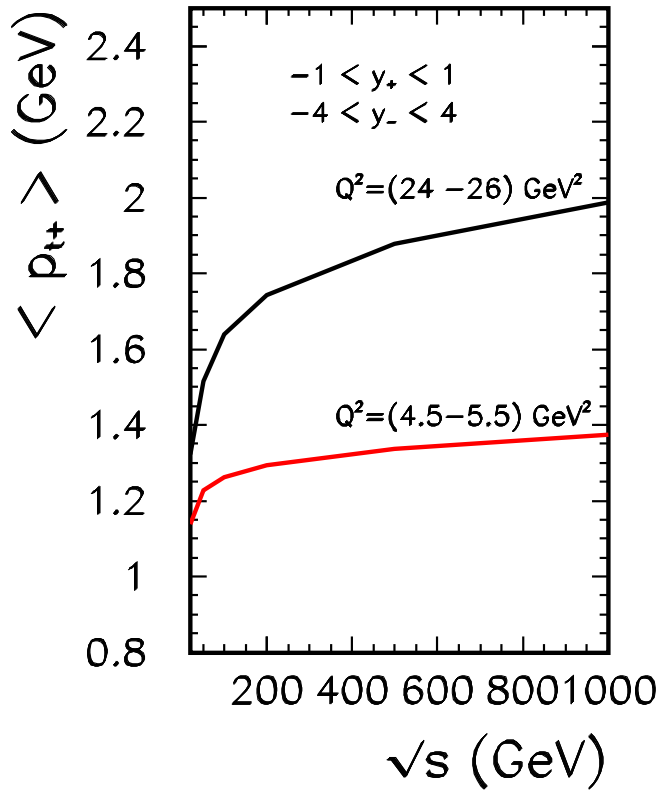}
\end{center}
   \caption{\label{fig:pt_plus_ave}
   \small
Average value of $p_{t+}$ as a function of center-of-mass energy
for two different windows of $Q^2$. Here $\mu_F^2 = Q^2$.
}
\end{figure}


We show the effect of broadening
also for the RHIC energy $\sqrt{s}$ = 200 GeV where we show distribution
in $p_{t+}$ for different bins in photon virtuality.
The effect of broadening is inherently related to the Kwieci\'nski UPDFs
where the initial $k_{1t}^2$ and/or $k_{2t}^2$ smearing depends
on the factorization scale and on $x_1$ and/or $x_2$ (see discussion in 
\cite{Kwiecinski,SRS07,PS07}).


\begin{figure}
\begin{center}
\includegraphics[width=0.6\textwidth]{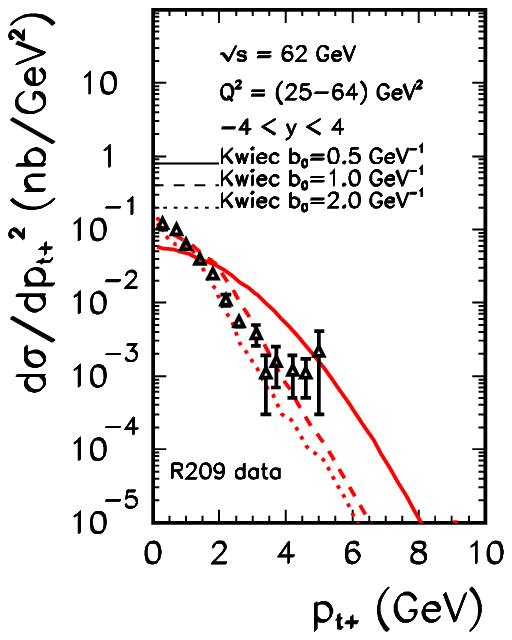}
\end{center}
   \caption{\label{fig:dsig_dptp2_r209_0-order}
   \small
Distribution in $p_{t+}$ for zero-order Drell-Yan in proton-proton
collisions for W = 62 GeV. Different curves correspond to different values
of the $b_0$ parameter in the Kwieci\'nski UPDFs.
The experimental data of the R209 collaboration are taken from
\cite{R209}.
}
\end{figure}


Finally we wish to confront our calculation with existing data
for the Drell-Yan dilepton production. In 
Fig.\ref{fig:dsig_dptp2_r209_0-order}
we show our results with different values of the parameter $b_0$
in the Kwieci\'nski UPDFs \cite{Kwiecinski}. The parameter $b_0$
quantifies nonperturbative effects which are beyond evolution
effects embodied in the Kwieci\'nski evolution equations. The dominant 
effect is probably the Fermi motion of nucleon constituents. This effect
is often neglected in the standard QCD calculations.
We get quite good description of the R209 collaboration data already
in the zeroth-order with $b_0$ = 1 - 2 GeV$^{-1}$. 
This is a bit surprising, as the zeroth-order contribution
is usually neglected in orthodox collinear approach. 
It seems therefore indespensible to include higher-orders 
in the $k_t$-factorization approach.
We shall discuss this issue in the next section.

\subsection{1-st order component with transverse momenta}

Let us now discuss contribution of processes of one order higher
than in the previous section, with hard subprocesses
shown in Fig.\ref{fig:drell_yan_1st_order}. These diagrams have
to be inserted between two partonic ladders.

The correlation in azimuthal angle between jet and the dilepton pair
is shown in Fig.\ref{fig:dsig_dphid_jet_lplm}. One can see
a strong deviation from the collinear back-to-back kinematics
caused by the transverse momenta of initial partons.
The most top (thick) lines are for the full phase space. The 
intermediate lines are with extra cuts on $e^+ e^-$ and jet rapidities.
The most bottom lines are for extra cuts on transverse momenta
of the associated jet $p_t(jet) >$ 5 GeV.
The latter cut changes drastically the shape of distributions
which are peaked now more at $\phi$ = 180$^0$. 


\begin{figure}
\begin{center}
\includegraphics[width=0.6\textwidth]{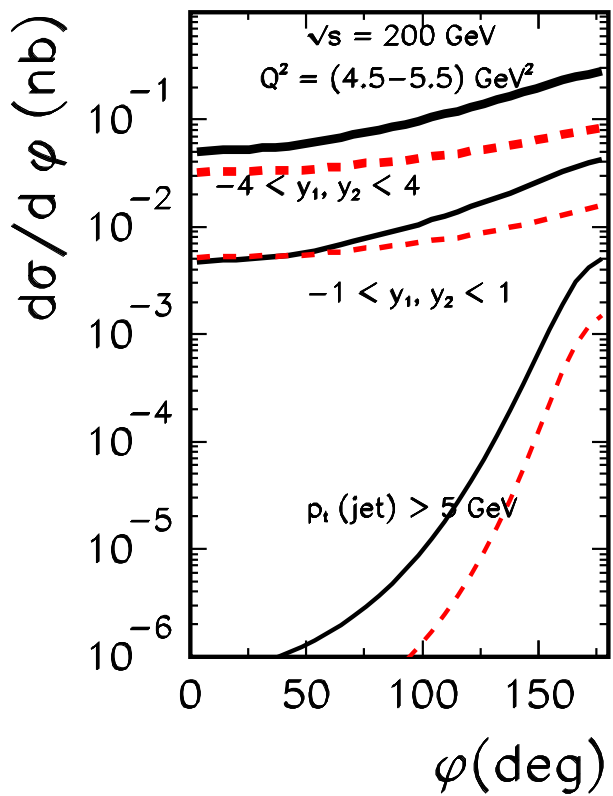}
\end{center}
   \caption{\label{fig:dsig_dphid_jet_lplm}
   \small
Distribution in azimuthal angle between jet and dilepton pair in proton-proton
collisions for the RHIC energy $\sqrt{s}$ = 200 GeV for the QCD Compton 
(solid lines)
and quark-antiquark annihilation (dashed lines). The results for 
-1 $< y(e^+ e^-), y(jet) <$ 1 are shown with thin lines and the results for 
-4 $< y(e^+ e^-), y(jet) <$ 4 are shown with thick lines. 
}
\end{figure}


In Fig.\ref{fig:dsig_dptp_0vs1-order} we show corresponding
transverse momentum distribution of the dilepton pair
for RHIC energy $\sqrt{s}$ = 200 GeV.
For comparison we show also zeroth-order contribution
discussed in the previous section. The zeroth-order contribution
dominates at small transverse momenta, while the first-order
contribution at transverse momenta larger than about 5 GeV.
Here we include both Compton and annihilation processes.
The first-order contributions have no singularity at $p_t(e^+ e^-)$
as in the collinear approach.
The zeroth and first-order contributions overlap only in a very limited 
range of $p_t(e^+ e^-) \approx$ 4 GeV.
The zeroth-order $k_t$-factorization component includes some higher
order contributions via UPDFs. 
However, because the two contributions occupy rather different 
parts of the phase space no severe double counting is expected.


\begin{figure}
\begin{center}
\includegraphics[width=0.6\textwidth]{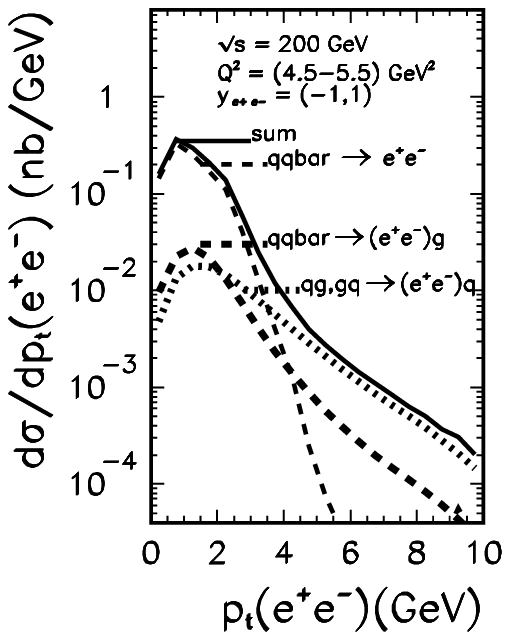}
\end{center}
   \caption{\label{fig:dsig_dptp_0vs1-order}
   \small
Distribution in $p_{t+}$ for the RHIC energy $\sqrt{s}$ = 200 GeV.
We show separately zero-order (dashed line) and first-order Compton
(dotted line) contributions. Here -1 $< y(e^+ e^-) <$ 1 and rapidity
of the jet for the first-order contribution is in the full phase space.
}
\end{figure}


When integrated the first-order contribution is significantly smaller
than the zeroth-order one. This can be better seen in 
Fig.\ref{fig:dsig_dy_0vs1-order} where we show rapidity distributions.
We show the zeroth-order (LO) and two $k_t$-factorization
Compton components: QCD Compton ($qg \to q e^+ e^-$ and $gq \to q e^+ e^-$)
and quark-antiquark annihilation ($q\bar q \to g e^+ e^-$ and $\bar q q \to g e^+ e^-$).


\begin{figure}
\begin{center}
\includegraphics[width=0.6\textwidth]{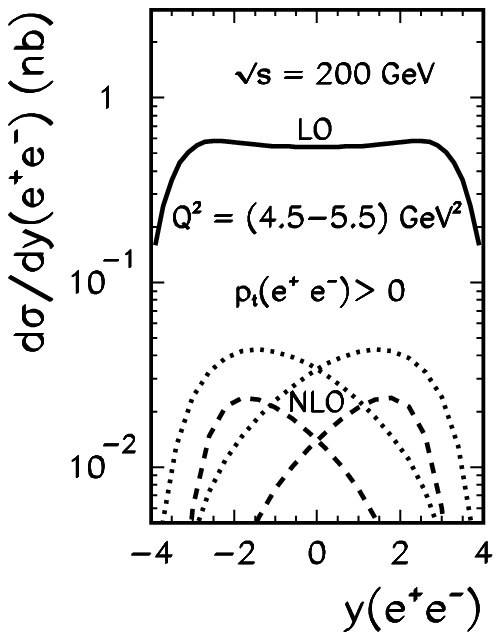}
\end{center}
   \caption{\label{fig:dsig_dy_0vs1-order}
   \small
Distribution in $y(e^+ e^-)$ for the RHIC energy $\sqrt{s}$ = 200 GeV.
We show separately zero-order (solid line) and first-order QCD Compton
(dotted line) and quark-antiquark annihilation (dashed line) contributions. 
The rapidity of the associated jet for the first-order contributions is integrated 
in the full phase space.}
\end{figure}


Let us present now the first-order contributions together with existing
data. In Fig.\ref{fig:dsig_dptp_0th_vs_1st_order_r209} we present separately
two first-order contributions: Compton and quark-antiquark annihilation.
The sum of the both contributions is shown by the thin solid line.
This contribution is about factor of 4 smaller than the R209 collaboration data.
For comparison we show also the zeroth-order contribution.
The 0th-order contribution is much larger than the 1st-order one.
The situation may change at larger transverse momenta.
The sum of the 0th- and 1st-order well describe the transverse momentum data.
The scenario discussed here seems quite different than the one presented
in the text book by Field \cite{Field_book} where the data were discribed
by a convolution of the collinear first-order component and some
extra, somewhat arbitrary, Gaussian smearing and where the zeroth-order 
contribution was completely ignored.

\begin{figure}
\begin{center}
\includegraphics[width=0.5\textwidth]{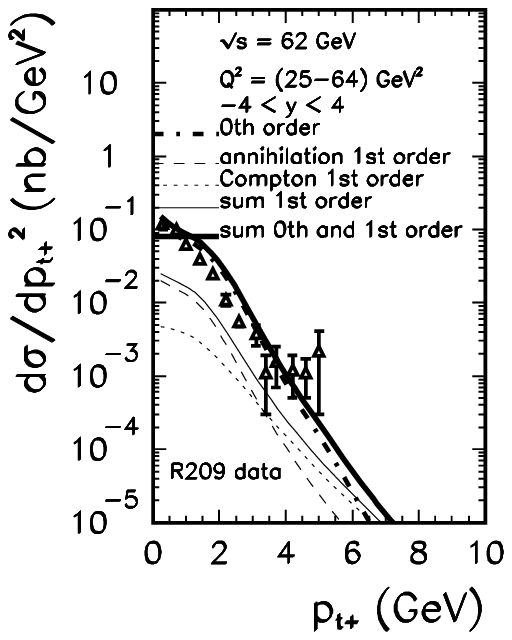}
\end{center}
   \caption{\label{fig:dsig_dptp_0th_vs_1st_order_r209}             
   \small Distribution in transverse momentum of the dilepton pair
in proton-proton collisions.
We show both first order contributions: QCD Compton (dotted line) and
quark-antiquark annihilation (dashed line). For comparison we show
also the zeroth order contribution (dash-dotted line).
}
\end{figure}


In Fig.\ref{fig:dsig_dptp_coll_kw} we compare distributions obtained
in collinear and $k_t$-factorization approach. The difference can be 
seen in small transverse momentum region. Above $p_{t+} >$ 4 GeV the two
approaches practically give the same results. The singularities seen
for collinear approximation disappear when transverse momenta of
initial partons are included.

\begin{figure}
\begin{center}
\includegraphics[width=0.5\textwidth]{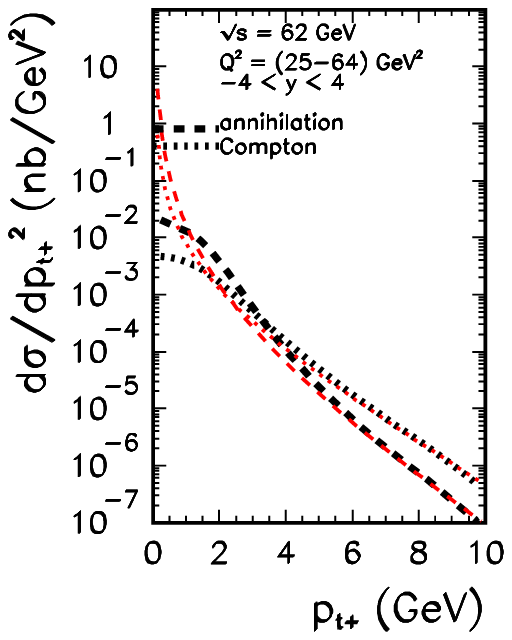}
\end{center}
   \caption{\label{fig:dsig_dptp_coll_kw}             
   \small Distribution in transverse momentum of the dilepton pair
for the R209 collaboration experiment.
The collinear (thin lines) distributions are compared to 
$k_t$-factorization (thick lines) distributions for the QCD Compton (dotted)
and quark-antiquark annihilation (dashed).
}
\end{figure}


The first-order contribution becomes dominant at larger transverse
momenta of the dilepton pairs. This is shown in 
Fig.\ref{fig:dsig_dptsum_ua1_scales}
where we present transverse momentum distribution of dimuons of opposite
charge for proton-antiproton scattering at the center-of-mass energy
$\sqrt{s}$ = 630 GeV. We confront results of our first-order 
$k_t$-factorization calculation against UA1 collaboration experimental 
data \cite{UA1} which were
measured at $p_{t,\mu\mu} >$ 10 GeV. We show our results for two different
choices of the factorization scale: (a) $\mu_F^2 = M_{\mu \mu}^2$ (left panel),
(b) $\mu_F^2 = M_{\mu \mu}^2 + p_{t,\mu \mu}^2$ (right panel). 
The experimental data are well described within theoretical uncertainties. 
In this case the zeroth-order contribution (not shown in the figure) 
is concentrated at $p_{t,\mu \mu} <$ 10 GeV.


\begin{figure}
\begin{center}
\includegraphics[width=0.4\textwidth]{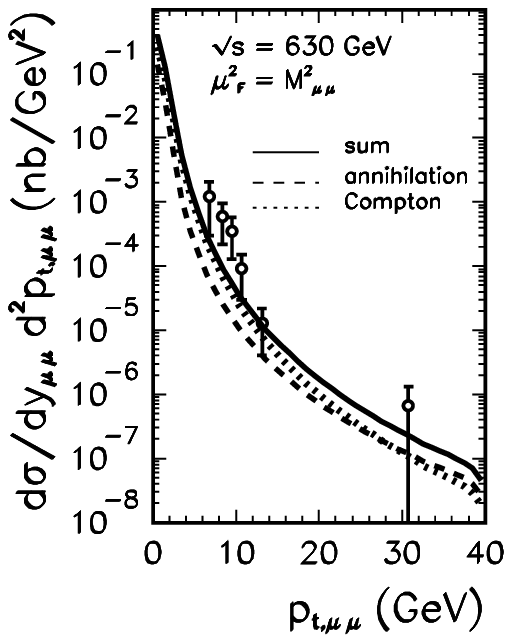}
\includegraphics[width=0.4\textwidth]{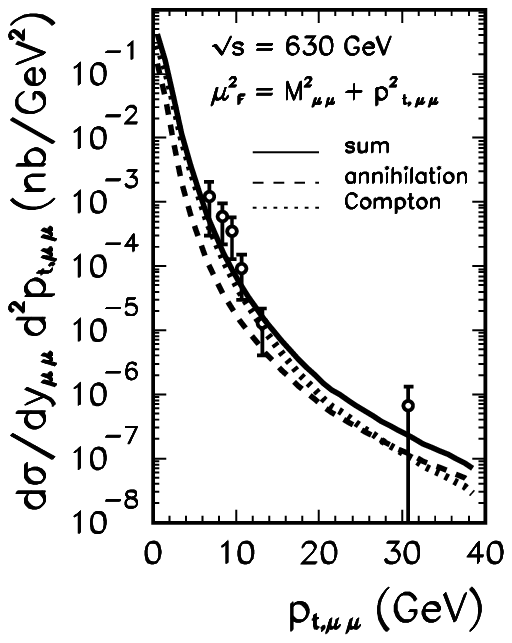}
\end{center}
   \caption{\label{fig:dsig_dptsum_ua1_scales}
   \small
Dilepton invariant mass distribution for first-order Drell-Yan 
processes in proton-antiproton scattering for two different scales:
$\mu_F^2 = Q^2$ (left panel) and $\mu_F^2 = Q^2 + p_{t,\mu \mu}^2$
(right panel) at $\sqrt{s}$ = 630 GeV. 
Here $Q^2 \in$ (1,2.5$^2$) GeV$^2$ and $y_2 \in$ (-1.7,1.7).
The contributions of Compton (dotted) and annihilation (dashed)
are shown separately. The experimental data of UA1 collaboration 
are taken from \cite{UA1}.   
}
\end{figure}


In the first-order collinear calculations the transverse momentum
of the dilepton pair is completely balanced by the transverse momentum
of the associated jet (quark or antiquark for the Compton process
and gluon for the quark-antiquark annihilation). This strict balance
is not longer true when transverse momenta of initial partons
are taken into account. In fact the disbalance can be a measure of
the transverse momenta of the intial partons.
In Fig.\ref{fig:map_pt1pt2} we show two-dimensional 
distributions
in ($p_{1t}(jet), p_{2t}(e^+ e^-)$) for Compton (upper panels)
and annihilation (lower panels) contributions
for $\sqrt{s}$ = 200 GeV and a 
narrow window in the photon virtuality. 
We see broad distributions of the strength along diagonal $p_{1t} = p_{2t}$
with a smearing of the order of a few GeV. This smearing is a consequence
of the convolution of two unintegrated parton distributions embodied
in Eq.(\ref{1st_order_kt-factorization_Compton}) and 
(\ref{1st_order_kt-factorization_annihilation}).
The broadening strongly depends on the choice of the factorization
scale (compare left and right panels).


\begin{figure}
\begin{center}
\includegraphics[width=0.4\textwidth]{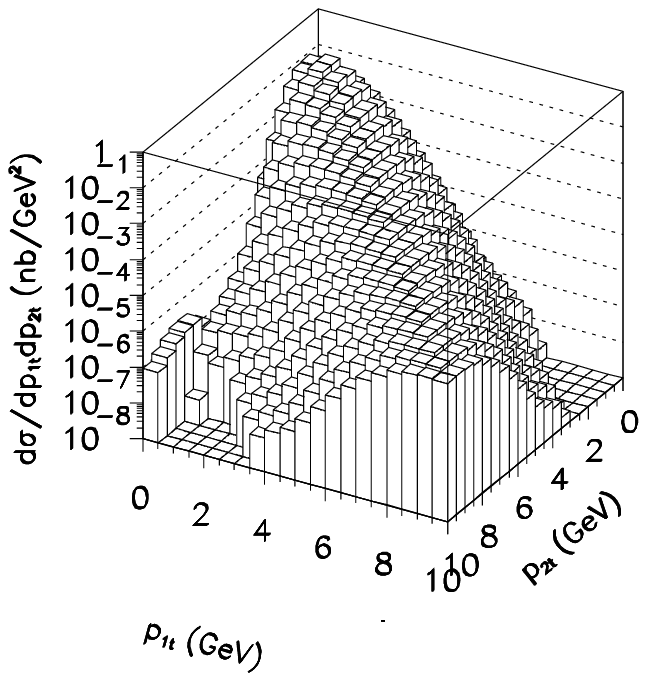}
\includegraphics[width=0.4\textwidth]{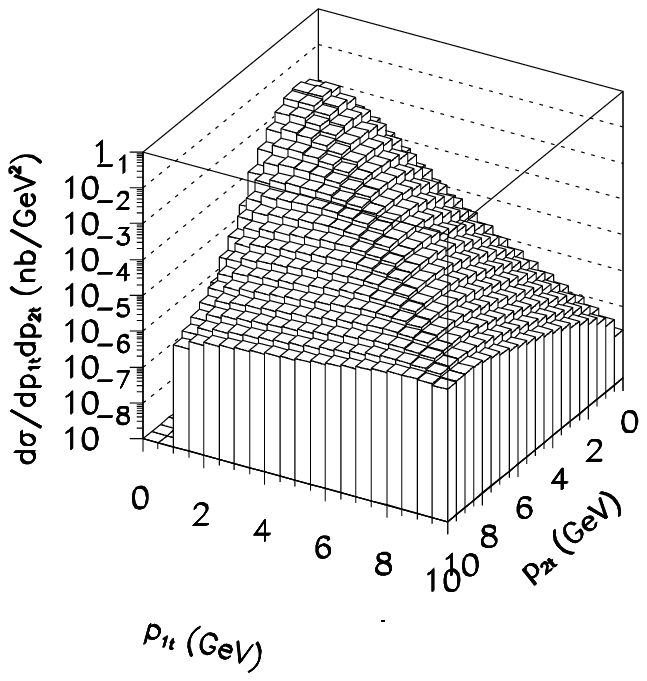}
\includegraphics[width=0.4\textwidth]{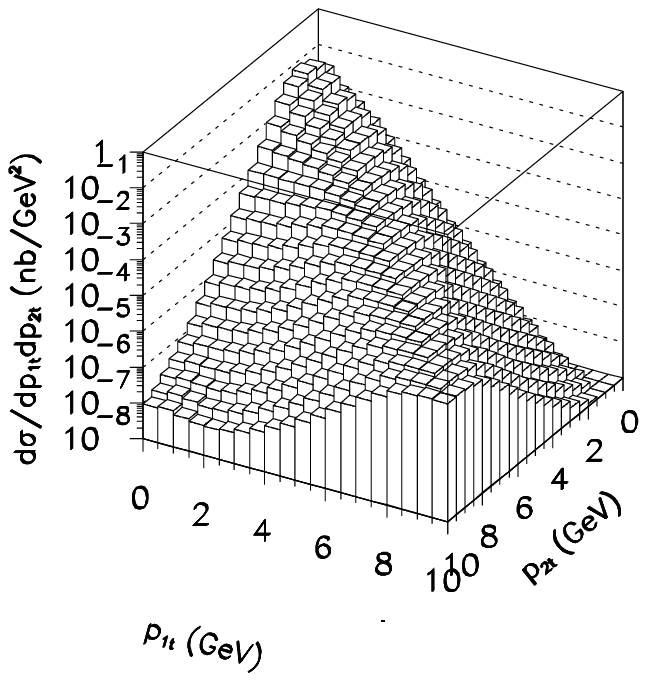}
\includegraphics[width=0.4\textwidth]{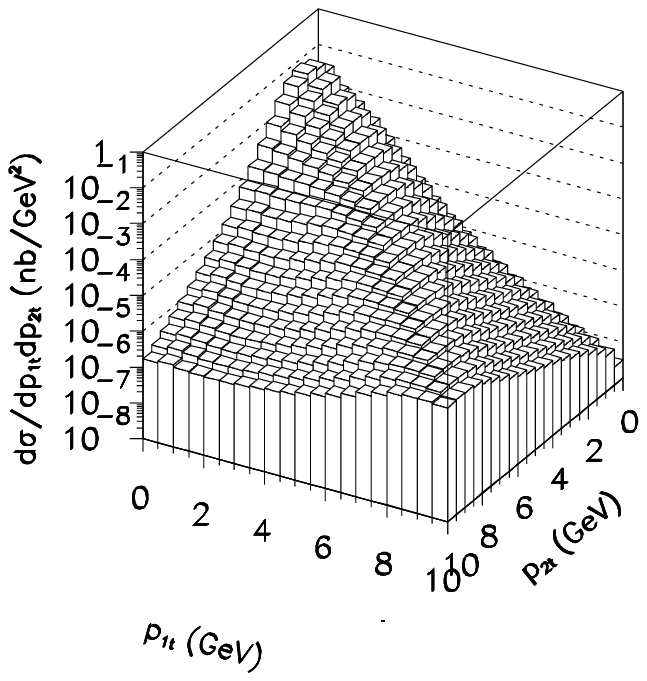}
\end{center}
   \caption{\label{fig:map_pt1pt2}
   \small
Two-dimensional distributions in $p_{1t}(jet)$ and $p_{2t}(e^+ e^-)$
for the first-order Compton contributions (upper panels)
and for the first-order quark-antiquark annihilation contributions 
(lower panels) in proton-proton collisions at $\sqrt{s}$ = 200 GeV
and $Q^2 \in$ (4.5,5.5) GeV$^2$. 
Left panels are for $\mu_F^2 = Q^2$ and right panels for 
$\mu_F^2 = Q^2 + p_{t,ee}^2$. No cuts on rapidity 
of the jet as well on the rapidity of the dilepton pair 
were applied.
}
\end{figure}





In Fig.\ref{fig:dsig_dp1tdp2t_dy_ua1} we show similar distributions for 
the UA1 collaboration experiment. Here the range of transverse momenta 
of the jet and dilepton pair is much larger. 
For such a broad range of transverse momenta it looks as if the cross section
is concentrated on the diagonal $p_{1t}(jet) = p_{2t}(\mu^+ \mu^-)$.
A closer inspection shows a smearing similar as shown previously
for proton-proton scattering at $\sqrt{s}$ = 200 GeV 
(see Fig.\ref{fig:map_pt1pt2}). 


\begin{figure}
\begin{center}
\includegraphics[width=0.4\textwidth]{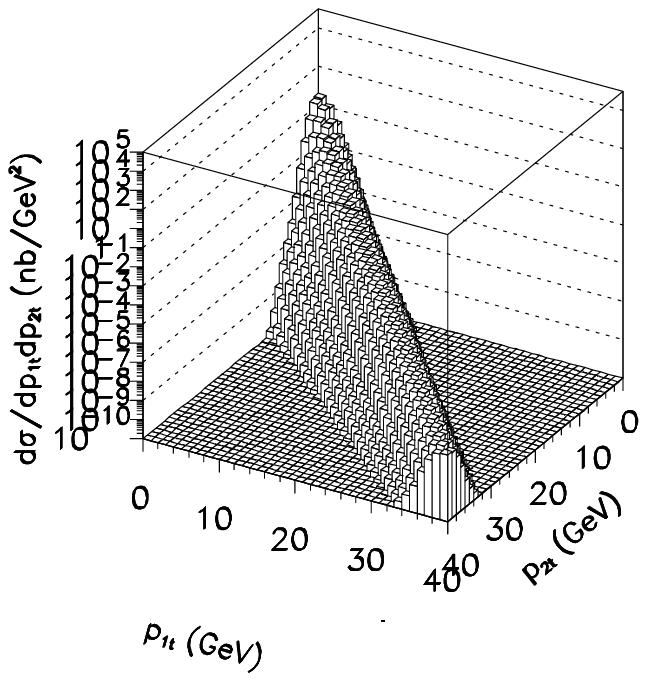}
\includegraphics[width=0.4\textwidth]{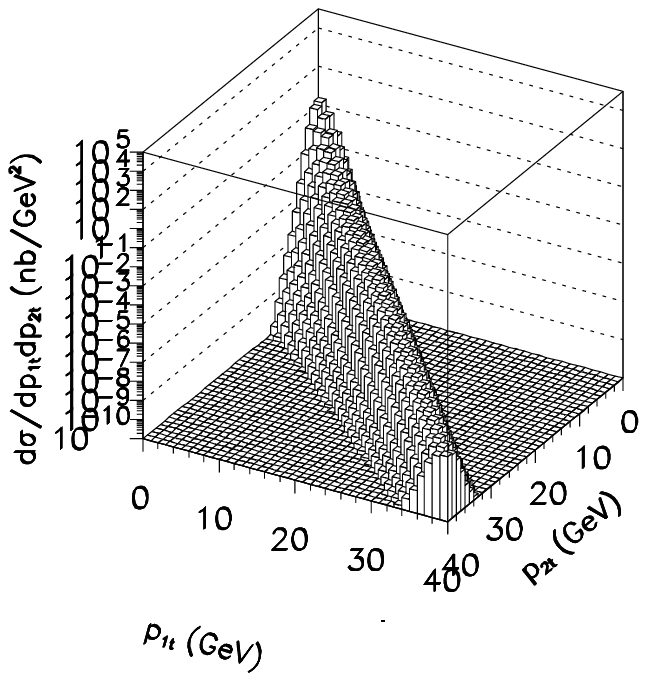}
\end{center}
   \caption{\label{fig:dsig_dp1tdp2t_dy_ua1}
   \small
$(p_{1t}(jet),p_{2t}(\mu \mu))$ distribution for first-order Drell-Yan 
processes in proton-antiproton collisions at $\sqrt{s}$ = 630 GeV.
Left panel for QCD Compton and the right panel 
for annihilation.   
Here $\mu_F^2 = Q^2$ and $Q^2$ = (1,2.5$^2$) GeV$^2$ and 
-1.7 $ < y(\mu^+ \mu^-) < $ 1.7.   
}
\end{figure}

\section{Conclusions}

We have calculated both zeroth- and first-order contributions 
to dilepton production in the formalism with transverse
momenta of initial partons taken into account. 
In these calculations we have used Kwieci\'nski unintegrated parton 
distributions which include both smearing in parton momentum 
due to nonperturbative effects in hadrons before collision as well as 
extra smearing due to QCD evolution effects in the collision process 
as encoded in the Kwieci\'nski evolution equations.

We have calculated correlations in azimuthal angle
between both charged leptons as well as correlations in the two-dimensional
space of transverse momentum of the positron and transverse momentum
of the electron. Both effect of the Fermi motion and effect of subsequent
emissions from the ladder lead to deviations from the delta function
in relative azimuthal angle centered at $\phi = \pi$ (collinear case)
and deviations from $p_t$(electron) = $p_t$(positron) condition.
The shape of the distribution in transverse momentum of the pair depends
both on incident energy and virtuality of the time-like photon.
This is a strightforward consequence of the QCD evolution encoded in the
Kwieci\'nski equations. We predict larger smearing in transverse momentum
of the dilepton pair for larger dilepton masses. The existing 
experimental data at $\sqrt{s}$ = 62 GeV can be well explained by 
the zero-order component by adjusting the parameter responsible for
nonperturbative effects of internal motion of partons in hadrons. 
This is rather in odds with the explanations in the literature, where 
the data are explained by an extra convolution of the first-order 
contribution with a Gaussian  smearing function.
In orthodox collinear the zero-order contribution is
completely ignored. The smeared zeroth-order contribution discussed here may 
be partly responsible for missing strength in the spectrum of so-called
nonphotonic electrons \cite{LMS08}.

We have also calculated dilepton transverse momentum distribution
in the first order for the matrix element. Inclusion of 
initial transverse momenta removes singularity at $p_{t,l^+ l^-}$ = 0.
The first-order contribution dominates only at larger transverse
momenta of the pair and is smaller than the zeroth-order contribution
at low transverse momenta. The inclusion of initial transverse momenta
leads naturally to decorellation of relative azimuthal angle of a jet and
dilepton pair (in the first-order collinear approximation they are emitted 
back-to-back). We have also discussed analogous decorrelations on the
($p_t(jet),p_t(l^+ l^-)$) plane. The initial transverse momenta lead
to sizeable deviations from the collinear condition $p_t(jet) = p_t(l^+ l^-)$.  
 
Finally we wish to make a comment on possible double counting.
In principle, our leading order contribution contains
diagrams which look like first order diagrams.
The standard pQCD first-order result contains large 
$log(Q^2/p_t^2)$. For fixed
$Q^2$ it happens when $p_t^2$ is small.
Fortunately it happens numerically that then 
(in our approach) the first-order result is much lower 
than the almost purely nonperturbative origin zeroth-order
result. In principle, the double counting
happens when the zeroth- and first-order results
are comparable. As shown in our calculation this
happens in a very narrow interval of $p_{t+}$.
So we expect rather small double counting.


\end{document}